\documentclass[aip,jcp,reprint,superscriptaddress,noshowkeys]{revtex4-2}
\usepackage{graphicx,dcolumn,bm,microtype,multirow,amscd,amsmath,amssymb,amsfonts,mathtools,physics,wrapfig,siunitx}

\usepackage[version=4]{mhchem}
\usepackage[dvipsnames]{xcolor}
\usepackage[normalem]{ulem}
\usepackage[utf8]{inputenc}

\usepackage{hyperref}
\hypersetup{
    colorlinks,
    linkcolor={red!50!black},
    citecolor={red!70!black},
    urlcolor={red!80!black}
  }

\usepackage{xspace}
\newcommand{\mc}{\multicolumn}

\newcommand{\mcc}[1]{\multicolumn{1}{c}{#1}}

\newcommand{\SupInf}{\textcolor{blue}{Supporting Information}\xspace}

\newcommand{\zz}{\phantom{(1)}}

\newcommand{\m}{\phantom{-}}

\usepackage[normalem]{ulem}
\newcommand{\titou}[1]{\textcolor{black}{#1}}

\newcommand{\LCPQ}{Laboratoire de Chimie et Physique Quantiques (UMR 5626), Universit\'e de Toulouse, CNRS, Toulouse 31062, France}

\makeatletter

\begin{document}	

\title{Reference Energies for Non-Relativistic Core Ionization Potentials}

\author{Antoine \surname{Marie}}
       \email{amarie@irsamc.ups-tlse.fr}
       \affiliation{\LCPQ}

       \author{Loris \surname{Burth}}
       \affiliation{\LCPQ}
       
\author{Pierre-Fran\c{c}ois \surname{Loos}}
       \email{loos@irsamc.ups-tlse.fr}
       \affiliation{\LCPQ}

\begin{abstract}     
Deep-lying core electrons carry highly localized, site-specific information that forms the basis of X-ray photoelectron spectroscopy.
Accurately predicting their associated core ionization potentials (IPs) is a demanding theoretical task, requiring a balanced treatment of strong orbital relaxation, electron correlation, and relativistic effects.
Over the years, a variety of approaches have been developed, ranging from state-specific wave function methods to linear-response formalisms and Green's function techniques. 
However, their assessment has often relied on comparisons with experiment, where multiple sources of error (basis set incompleteness, relativistic corrections, and vibrational effects) are entangled, making it difficult to isolate the performance of correlation treatments.     
In the present work, we establish a consistent, theory-based benchmark for core IPs by computing 84 non-relativistic values (73 second-row and 11 third-row IPs) at the full configuration interaction level within the core-valence separation approximation, using large correlation-consistent basis sets augmented with tight-core and diffuse functions (aug-cc-pCVXZ).
These results define theoretical best estimates within a fixed finite basis set, providing a chemically accurate reference for method development and validation. 
Importantly, our dataset allows for systematic, theory-versus-theory comparisons that disentangle correlation and relaxation effects from other physical contributions. 
On this basis, we assess the performance of widely used approximate methods, including equation-of-motion coupled-cluster approaches up to the inclusion of quadruple excitations, the one-shot $G_0W_0$ scheme, as well as state-specific methods.
 \bigskip
 \begin{center}
 	\boxed{\includegraphics[width=0.5\linewidth]{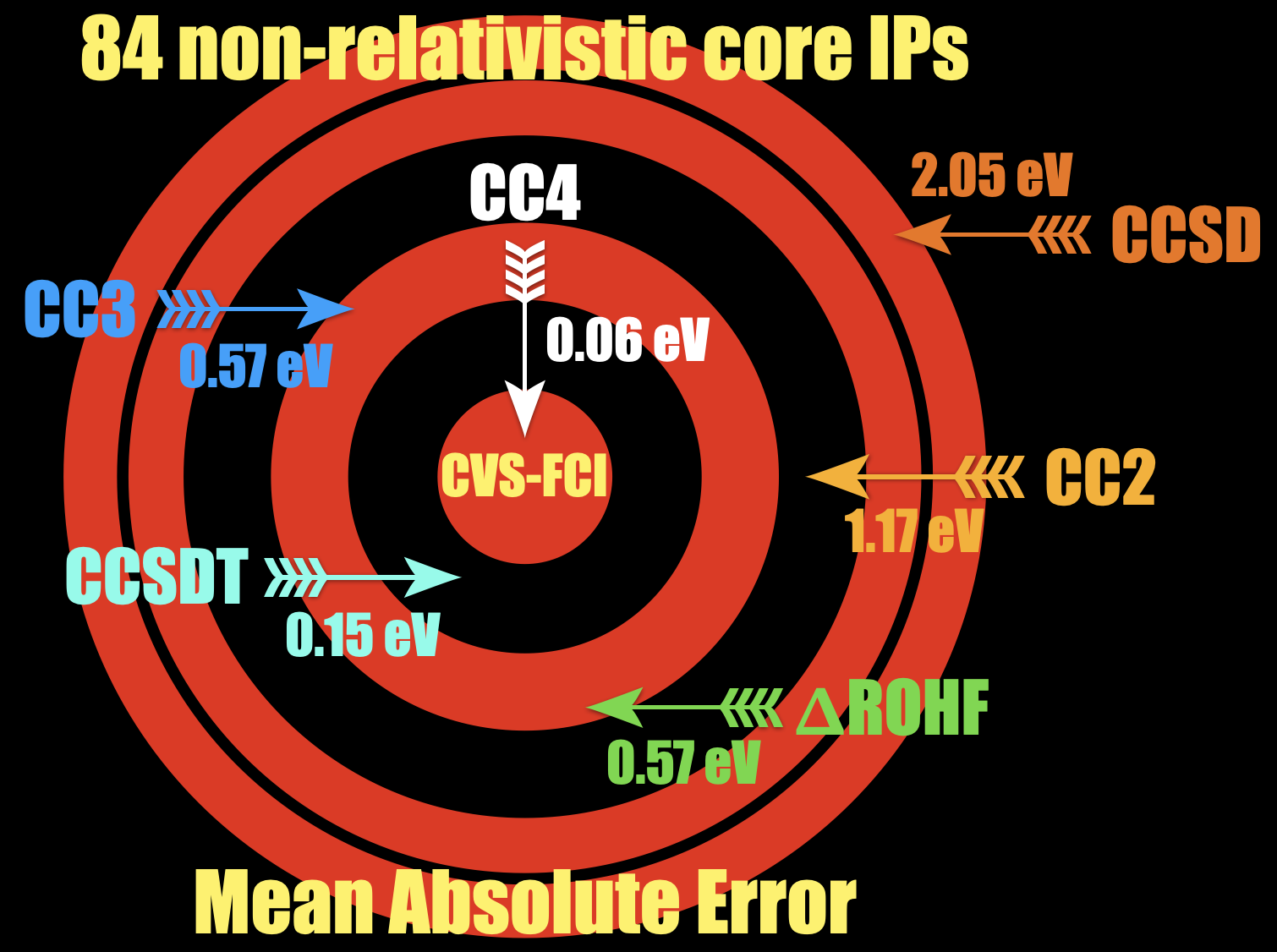}}
 \end{center}
 \bigskip
\end{abstract}

\maketitle

\section{Introduction}
\label{sec:intro}

In molecular systems, core-ionized states arise when a deeply bound core electron is removed, most commonly through photoionization. 
These electrons are highly energetic and spatially localized, resulting in a positively charged molecular ion with a core vacancy. 
Due to the atomic specificity of core electrons, the ionization process is inherently atom-selective.
In addition, their subtle dependence on the surrounding environment enables X-ray photoelectron spectroscopy (XPS) to distinguish between chemically inequivalent atoms of the same element.\cite{Nordling_1957,Sokolowski_1958,Siegbahn_1973,Fadley_2010,Greczynski_2020,Stevie_2020,Greczynski_2022}
This makes XPS a powerful probe of local chemical environments within complex molecules.  

From a theoretical point of view, the accurate description of core-ionized states requires a subtle balance between many effects, which can be challenging for conventional electronic structure methods.\cite{Bagus_2013,Norman_2018}
Firstly, the difficulty of accounting for significant relaxation effects is added to the usual problem of accurately treating electronic correlation.
The sudden removal of a core electron induces a strong local perturbation in the electronic structure, triggering a reorganization of the valence electron density.
This phenomenon must be properly described to obtain reliable results.\cite{Bagus_1965,Bagus_1972,Cederbaum_1977a}
Such effects necessitate the use of basis sets with increased flexibility in the core region, typically achieved by including additional tight-core functions.\cite{Carniato_2002,Besley_2009,Ambroise_2019,Sarangi_2020,Ambroise_2021,Delgado_2025}
In addition, core electrons are subject to much stronger relativistic effects than valence electrons, which must be accounted for to achieve quantitative prediction of core ionization potentials (IPs).\cite{Niskanen_2011,Keller_2020,Opoku_2022,Kehry_2023,Halbert_2025}
For second-row elements, scalar relativistic corrections are usually sufficient, while for third-row atoms, higher-level corrections such as spin-orbit coupling must be included.\cite{Zheng_2022}

While the aforementioned effects are the primary contributors to core-electron binding energies, several additional effects must be accounted for to faithfully describe core-ionized states.
In particular, these states are resonances, i.e., they are embedded in the continuum,\cite{Jagau_2022} and are thus meta-stable, capable of decaying non-radiatively via the ejection of a valence electron.
This process, known as the Auger-Meitner decay, requires a treatment of continuum states, either through explicit or implicit approaches.\cite{Meitner_1922,Auger_1923,Matz_2022,Matz_2023,Jayadev_2023,Ferino-Perez_2024}
Finally, the current resolution of XPS is such that they can be vibrationally resolved. 
It thus necessitates the inclusion of vibrational effects, and relaxation of the geometry of the core-ionized state, in any comprehensive theoretical description of XPS.\cite{Borve_2000,Sankari_2003,Thiel_2003,Hatamoto_2007,Huang_2024,Alvertis_2024}
Note that the prediction of condensed-phase XPS, in particular for surface characterization, introduces further complexity beyond the challenges encountered in the gas phase.\cite{Aoki_2018,Ishii_2010,Zhu_2021,Kahk_2023,Mukatayev_2024}

The core IPs of second-row atoms typically lie in the range of several hundred \si{\eV}, and can exceed a thousand \si{\eV} for third-row atoms. 
Although these absolute values are large, core-level binding energies exhibit subtle variations depending on the chemical environment of the ionized atom.
These so-called chemical shifts are key to interpreting experimental spectra,\cite{Bagus_2013} and theoretical prediction of XPS spectra and chemical shifts is a valuable, if not mandatory, tool for experimental interpretation.
For instance, resolving different chemical environments in carbon 1s XPS spectra requires a theoretical accuracy of at least \SI{0.1}{\eV}.
Given the challenges outlined above, reaching such precision in computed core IPs hinges on a delicate interplay of error cancellation and accumulation.

The first theoretical study of core-ionized molecules was reported by Bagus in 1965.\cite{Bagus_1965}
In this seminal work, the core IPs were computed as the energy difference between two Hartree-Fock (HF) calculations: one where the orbitals were optimized for the neutral ground-state determinant, and the other for the core-ionized, non-Aufbau determinant.
This state-specific approach represented a significant improvement over Koopmans' theorem, which estimates IPs directly from orbital energies without accounting for orbital relaxation.\cite{SzaboBook,PueyoBellafont_2015,Hirao_2022}
The success of Bagus' method lies in its explicit inclusion of relaxation effects via separate optimization of the core-ionized state.
Since then, the state-specific formalism has been extended to various methods such as general mean-field approaches,\cite{Cavigliasso_1999,Takahata_2003,Besley_2009,Holme_2011,Besley_2012,PueyoBellafont_2015,PueyoBellafont_2016a,PueyoBellafont_2017,Vines_2018} M{\o}ller-Plesset perturbation theory,\cite{Holme_2011,Smiga_2018,Morgunov_2024} coupled-cluster (CC) theory,\cite{Morgunov_2024,Holme_2011,Zheng_2019,CFOUR,Zheng_2022,Arias-Martinez_2022} symmetry-adapted-cluster configuration interaction (SAC-CI),\cite{Ohtsuka_2006} complete-active-space self-consistent field,\cite{Bagus_2016} and the driven similarity renormalization group approach.\cite{Huang_2024,Huang_2025}

The linear-response formalism offers an alternative route for computing core IPs, wherein they are calculated on top of a ground-state reference wave function, such as in equation-of-motion (EOM) CC,\cite{Besley_2012,Coriani_2015,Vidal_2019,Liu_2019,CFOUR,Matthews_2020a,Jangid_2024,Manna_2024} or SAC-CI.\cite{Sankari_2003,Kuramoto_2005,Hatamoto_2007}
In these methods, core IPs appear as high-lying eigenvalues of the linear-response eigenvalue problem.
Such states can be accessed either via root-homing techniques, which enable the direct targeting of inner-spectrum eigenvalues,\cite{Butscher_1976,Dorando_2007,Hillenbrand_2025} or, more commonly, through the core-valence separation (CVS) approximation introduced by Cederbaum and co-workers.\cite{Cederbaum_1980a}
The CVS approximation effectively decouples core-ionized states from those with a filled core shell, rendering the state of interest the ground state within the truncated CVS excitation space.
This simplification allows for the use of standard diagonalization algorithms such as Davidson's method.\cite{Davidson_1975}
Conceptually, CVS corresponds to considering only the bound part of the wave function and neglecting the coupling between the core-ionized state and the continuum.\cite{Liu_2019,Halbert_2021,Jagau_2022}

Alternatively, core IPs can also be computed as the poles of the one-body Green's function. \cite{Onida_2002,Martin_2016}
The $GW$ method, \cite{Hedin_1965,Reining_2018,Golze_2019,Marie_2024a} which has been highly successful in computing valence IPs of both molecules and solids, has recently been extended to the computation of core IPs.\cite{vanSetten_2018,Golze_2018,Golze_2020,Mejia-Rodriguez_2021,Li_2022,Mukatayev_2023}
This required significant numerical and methodological developments, but core IPs can now be routinely computed at the $GW$ level.\cite{Golze_2018,Duchemin_2020,Panades-Barrueta_2023}
The performance of others Green's function approximations, such as the particle-particle $T$-matrix self-energy~\cite{Zhang_2017,Li_2021b} or the algebraic-diagrammatic construction, has been evaluated for core IPs.\cite{Angonoa_1987,Angonoa_1989,Schirmer_2001,Thiel_2003,Kryzhevoi_2011,deMoura_2022,deMoura_2024}


The previous paragraphs show that a whole zoo of methods is now available for computing absolute core IPs and chemical shifts. 
This diversity calls for systematic benchmarking to assess the accuracy and limitations of each approach.
Takahata and Chong conducted an early benchmark comparing core IPs obtained from a state-specific formalism using 23 exchange-correlation functionals against 59 experimental reference values.\cite{Takahata_2003}
Later, Holme \textit{et al.}~further assessed the accuracy of numerous exchange-correlation functionals, as well as several wave function methods, across 77 carbon (1s) IPs.\cite{Holme_2011}
Pueyo-Bellafont and co-workers have then pursued this effort of benchmarking density-functional approximations with respect to an experimental set of values.\cite{PueyoBellafont_2016,PueyoBellafont_2016a}
The CORE65 benchmark data set, gathering 65 experimental core IPs of small to medium-sized molecules, has been used to assess the accuracy of various Green's function methods.\cite{Golze_2018,Golze_2020,Li_2021,Li_2022,Panades-Barrueta_2023}
However, comparisons with experimental data make it difficult to pinpoint the precise origin of the mean absolute error as contributions from correlation, relativistic effects, coupling with the continuum, basis set incompleteness, and vibrational effects are all entangled.
To overcome this, Liu and co-workers recently computed 21 core IPs at the CVS-IP-EOM-CCSDTQ level in various basis sets, allowing for a precise assessment of errors due to approximations in correlation and relaxation.\cite{Liu_2019}

The goal of the present work is to pursue this effort by designing a benchmark of 84 non-relativistic core IPs computed at the CVS full configuration interaction (CVS-FCI) level.\cite{Marie_2025}
All values are reported in the aug-cc-pCVTZ basis set at least, and in the aug-cc-pCVQZ basis set when computationally feasible.
To the best of our knowledge, the present dataset represents the first large-scale collection of CVS-FCI benchmark values for core IPs.
The primary objective of this work is to establish non-relativistic reference values in a fixed finite basis set, with the goal of assessing the accuracy of various approximate correlation methods.
Together with Ref.~\onlinecite{Marie_2024}, it naturally extends to charged excitations (i.e., excitations involving a change in the number of electrons) the \textsc{quest} database, a project initiated in 2018 with the goal of providing the quantum chemistry community with high-quality theoretical best estimates (TBEs) for vertical neutral excitation energies. \cite{Loos_2018a,Loos_2020d,Veril_2021,Loos_2025}
In this context, a TBE refers to a vertical excitation energy, computed with the best available \textit{ab initio} quantum chemical methods, that is believed to be chemically accurate, that is, within 1 kcal/mol (or \SI{0.043}{\eV}), with respect to the exact non-relativistic result in a given finite basis set.
The database now contains 1,489 vertical (neutral) excitation energies and has served as a reference for benchmarking a wide range of electronic structure methods (see Ref.~\onlinecite{Loos_2025}).
In the spirit of the \textsc{quest} database, \cite{Loos_2025} our focus is exclusively on theory-versus-theory comparisons, relying entirely on computational data without reference to experimental measurements.
A meaningful comparison with experiment would require, at the very least, corrections for basis set incompleteness and the inclusion of relativistic effects, considerations that lie beyond the scope of the present study.

\section{Computational details}
\label{sec:comp_det}

\subsection{Geometries and basis sets}

The ground-state geometries of the present set of molecules are taken from the \textsc{quest} database, \cite{Loos_2020d,Veril_2021,Loos_2025} or optimized following the same procedure, i.e., at the CC3/aug-cc-pVTZ level \cite{Christiansen_1995b,Koch_1997} without frozen-core approximation.
The geometry optimizations have been performed using \textsc{cfour}, \cite{CFOUR} and they have been collected in the \SupInf for the sake of completeness.

The present set of core IPs has been calculated using Dunning's aug-cc-pCVXZ (where X = D, T, and Q) correlation-consistent basis set augmented with tight-core and diffuse functions.\cite{Dunning_1989,Kendall_1992,Prascher_2011,Woon_1993,Peterson_2002}
For brevity, we refer to these basis sets as ACVXZ throughout the manuscript.
Note that the basis set description of the core-ionized state has been extensively studied, and there are more efficient hybrid schemes to converge to the complete basis set limit in practice.
We refer the reader to the recent work of Delgado and Matthews, and the bibliographical review therein, on this matter.\cite{Delgado_2025}
However, we decided to use a more standard basis set family for benchmarking purposes.

\subsection{Selected CI calculations}
\label{sec:sCI}

The FCI estimates reported in this study were obtained using selected configuration interaction (SCI) calculations performed with the CIPSI (Configuration Interaction using a Perturbative Selection made
Iteratively) algorithm,\cite{Huron_1973,Giner_2013,Giner_2015,Caffarel_2016b,Garniron_2017,Garniron_2018} as implemented in \textsc{quantum package}.\cite{Garniron_2019}

These reference core IPs were obtained in a state-specific fashion.
Ground-state calculations were carried out for the neutral molecule with all electrons correlated.
The core-ionized state was first optimized at the restricted open-shell HF (ROHF) level using the maximum overlap method (MOM)\cite{Gilbert_2008,Barca_2014,Barca_2018a} as described in Ref.~\onlinecite{Besley_2009}, and a CVS-SCI calculation was then performed in which the determinant space was restricted to configurations containing at least one hole in the targeted core orbital.\cite{Ferte_2023}
\titou{Note that there are some slightly different variants of the CVS approximation which also remove double core hole determinants from the CI expansion or that freeze the core electrons in the neutral ground-state calculation.
Here, a minimal CVS approximation has been used as the aim is to provide results as close as possible to FCI without any approximations.}

Variational energies were extrapolated to the FCI limit as a function of the second-order perturbative correction, using weighted linear fits over the last 3--6 CIPSI iterations.\cite{Holmes_2017,Damour_2021,Damour_2023,Burton_2024,Marie_2024}
Weights were defined as the square of the inverse second-order correction.
Among the set of extrapolated values obtained with different point selections, the one with the smallest standard error was retained as the FCI estimate.
Reported error bars correspond to these fitting uncertainties but do not represent statistical confidence intervals.
Extrapolations were carried out with \textsc{mathematica}.\cite{Mathematica}

To accelerate convergence, each CIPSI calculation was first truncated when the wave function exceeded $2\times10^6$ determinants.
Natural orbitals from this intermediate wave function were then constructed, and a second CIPSI calculation was performed in the natural orbital basis until convergence.
Note that CVS-FCI, due to its restricted space of determinants, is not independent with respect to the underlying set of orbitals.
This will be discussed in more detail in Sec.~\ref{sec:small-molecules}.

\subsection{Coupled-cluster calculations}
\label{sec:CC}

Core IPs were computed within the IP-EOM-CC framework using \textsc{cfour} with default convergence thresholds.\cite{CFOUR}
All electrons were correlated in the ground-state calculations, while the CVS approximation was applied for the EOM treatment of the core-ionized states.
\titou{The IP-EOM-CC calculations are performed using canonical RHF ground-state orbitals.}
Calculations were carried out at the CC2, \cite{Christiansen_1995a,Hattig_2000} CCSD, \cite{Purvis_1982,Scuseria_1987,Koch_1990a,Koch_1990c,Stanton_1993a,Stanton_1993b} CC3, \cite{Christiansen_1995b,Koch_1995,Koch_1997,Hald_2001,Paul_2021} CCSDT, \cite{Noga_1987,Scuseria_1988,Watts_1994,Kowalski_2001,Kowalski_2001a,Kucharski_2001} CC4, \cite{Kallay_2004b,Kallay_2005,Loos_2021a,Loos_2022b} and CCSDTQ \cite{Oliphant_1991,Kucharski_1991,Kallay_2001,Hirata_2004,Kallay_2003,Kallay_2004a} levels of theory.

Diagonalization of the CC effective Hamiltonian in the ($N-$1)-electron sector of Fock space \cite{Musial_2003a,Kamiya_2006,Gour_2006,Shavitt_2009} was carried out for CCSD, CCSDT, and CCSDTQ.
At the CCSD level, the EOM excitation manifold comprises one-hole (1h) and two-hole-one-particle (2h1p) configurations; three-hole-two-particle (3h2p) and four-hole-three-particle (4h3p) contributions are additionally included at the CCSDT and CCSDTQ levels, respectively.
As usual within the CC formalism, IP and electron affinity (EA) sectors are assumed to be decoupled. \cite{Nooijen_1995,Tolle_2023}

For CC2, CC3, and CC4, direct diagonalization in the $(N-1)$-electron sector is not yet available.
In these cases, core IPs were instead obtained from diagonalization in the $N$-electron sector \cite{Geertsen_1989,Stanton_1993a,Comeau_1993,Shavitt_2009} augmented with a diffuse ``bath'' orbital of zero energy.
At the CC2 level, the EOM space then consists of one-hole-one-particle (1h1p) and two-hole-two-particle (2h2p) configurations, while three-hole-three-particle (3h3p) and four-hole-four-particle (4h4p) excitations are added at the CC3 and CC4 levels, respectively.
Although both schemes yield identical IPs, the $N$-electron formulation is computationally more demanding because of the larger configuration space involved. \cite{Stanton_1999}
In each scheme, the desired states have been obtained thanks to the overlap-based root-following Davidson algorithm implemented in \textsc{cfour}.

\subsection{State-specific calculations}
\label{sec:SS}

The state-specific $\Delta$SCF calculations used a RHF reference for the neutral systems, and ROHF or UHF references for the core-ionized states.
They were carried out with \textsc{quantum package} \cite{Garniron_2019,Damour_2024a} and ORCA, \cite{Neese_2025} respectively.
As in the SCI calculations, the core-ionized states were optimized using MOM. \cite{Gilbert_2008,Besley_2009,Barca_2014,Barca_2018a}
The state-specific $\Delta$MP2 were computed on top of the unrestricted $\Delta$SCF using ORCA. \cite{Neese_2025}



\begin{table*}
  \caption{Core IPs (in \si{\eV}) of small molecules in the ACVTZ basis set computed at various levels of theory: CVS-FCI with optimized orbitals (``optimized'') or neutral HF orbitals (``Neutral''), CVS-IP-EOM-CC4 and -CCSDTQ computed in the ACVTZ basis set or based on a composite basis set scheme using CVS-IP-EOM-CCSDT and the ACVDZ basis set.
    \titou{The IP-EOM-CC calculations are performed using canonical RHF ground-state orbitals.}
	The extrapolation errors associated with the CVS-FCI values are reported in parentheses.
	Statistical descriptors of errors with respect to ``optimized'' CVS-FCI are also reported.}
  \label{tab:tab1}
    \begin{ruledtabular}
    \begin{tabular}{lcccccc}
      & \mc{2}{c}{CVS-FCI} & & & \mc{2}{c}{Composite}  \\
      \cline{2-3} \cline{6-7}
      \mcc{} & \mcc{optimized} & \mcc{neutral} & \mcc{CC4} & \mcc{CCSDTQ} & \mcc{CC4} & \mcc{CCSDTQ} \\
      \hline
      H\textbf{F$^*$}     & 694.10 & 694.10\zz & 694.02 & 693.99 & 694.12 & 694.09 \\
      H$_2$\textbf{O$^*$} & 539.87 & 539.86\zz & 539.78 & 539.77 & 539.87 & 539.86 \\
      \textbf{N$^*$}H$_3$ & 405.63 & 405.62(1) & 405.56 & 405.57 & 405.63 & 405.65 \\
      \textbf{C$^*$}H$_4$ & 290.85 & 290.85\zz & 290.81 & 290.82 & 290.86 & 290.88 \\
      B\textbf{F$^*$}     & 695.43 & 695.41\zz & 695.31 & 695.31 & 695.40 & 695.41 \\
      C\textbf{O$^*$}     & 542.47 & 542.44\zz & 542.44 & 542.34 & 542.53 & 542.43 \\
      \textbf{C$^*$}O     & 296.26 & 296.22\zz & 296.29 & 296.29 & 296.31 & 296.31 \\
      \textbf{N$^*_2$}    & 409.88 & 409.85\zz & 409.87 & 409.91 & 409.90 & 409.95 \\
      \textbf{F$_2^*$}    & 696.41 & 696.40\zz & 696.35 & 696.33 & 696.46 & 696.44 \\
      HC\textbf{N$^*$}    & 406.77 & 406.77\zz & 406.70 & 406.73 & 406.74 & 406.78 \\
      H\textbf{C$^*$}N    & 293.40 & 293.38\zz & 293.40 & 293.41 & 293.42 & 293.44 \\
      \hline
      MSE                 &        &  -0.02\zz &  -0.05 &  -0.06 & 0.06 & \m0.01 \\
      MAE                 &        & \m0.02\zz & \m0.06 & \m0.07 & 0.06 & \m0.03 \\
      RMSE                &        & \m0.02\zz & \m0.07 & \m0.08 & 0.07 & \m0.03 \\
      SDE                 &        & \m0.01\zz & \m0.04 & \m0.06 & 0.03 & \m0.03 \\
      Min                 &        &  -0.04\zz &  -0.12 &  -0.13 & 0.02 &  -0.03 \\
      Max                 &        & \m0.00\zz & \m0.03 & \m0.04 & 0.10 & \m0.06 \\
    \end{tabular}
  \end{ruledtabular}
\end{table*}

\subsection{$GW$ calculations}
\label{sec:GW}

The (one-shot) $G_0W_0$ calculations have been performed with PySCF \cite{pyscf} using the PBE functional with 45\% of exact exchange for second-row core IPs, as recommended in Ref.~\onlinecite{Golze_2020} for molecular systems, and 70\% of exact exchange for third-row core IPs (see discussion below).
The quasiparticle energies are obtained by solving the frequency-dependent quasiparticle equation (i.e., without linearization), and the broadening parameter $\eta$ was set to zero.

\section{Results}
\label{sec:results}

We employ standard statistical estimators in our analysis: mean signed error (MSE), mean absolute error (MAE), standard deviation of the errors (SDE), root-mean-squared error (RMSE), as well as the maximum and minimum errors (Max and Min).

\subsection{Small molecules}
\label{sec:small-molecules}

This study starts by focusing on 11 core IPs of small molecules for which the CVS-IP-EOM-CC4 and -CCSDTQ results could be computed in the ACVTZ basis set.
For conciseness, the CVS-IP-EOM prefix will be omitted in the following discussion.
\titou{The notation used for IPs is defined as follows: C\textbf{O$^*$} refers to the ionization of an electron from the oxygen $1s$ orbital of the \ce{CO} molecule.}
The corresponding values are reported in Table~\ref{tab:tab1} alongside two sets of CVS-FCI values corresponding to different sets of orbitals.
Indeed, once the CVS approximation is enforced, the invariance of the FCI energy with respect to orbital rotations is lost.
The first set of values, referred to as ``optimized orbitals'', serves as the reference in this work.
These were obtained by first generating a CVS-FCI expansion up to approximately two million determinants, built on top of a restricted open-shell determinant optimized to describe the core-ionized state.
Then, the natural orbitals corresponding to this CI wave function were computed, and a new CVS-FCI calculation was performed using these natural orbitals until convergence.

The second set of CVS-FCI results was obtained using the canonical orbitals of the neutral HF ground state, that is, the same orbitals as in IP-EOM-CC.
Table~\ref{tab:tab1} shows that the impact of orbital choice on the core IPs is minimal, with an average absolute deviation of only \SI{0.02}{eV}.
Moreover, the two CC methods that include quadruple excitations, whether approximately (CC4) or exactly (CCSDTQ), provide excellent agreement with the reference values, yielding MAE of \SI{0.06}{eV} and \SI{0.07}{eV}, respectively.

Finally, Table~\ref{tab:tab1} also reports CC4 and CCSDTQ results obtained using a composite basis set scheme, which has been shown to perform very well for neutral excitations. \cite{Silva-Junior_2010a,Loos_2018a,Loos_2019c,Loos_2020c,Loos_2022b,Jacquemin_2023,Loos_2024}
In this scheme, the core IPs are approximated as
\begin{multline} \label{eq:composite}
  E_{\text{CCSDTQ/ACVTZ}} \simeq E_{\text{CCSDTQ/ACVDZ}} \\
  + ( E_{\text{CCSDT/ACVTZ}} - E_{\text{CCSDT/ACVDZ}} ),
\end{multline}
where $E_{\text{method/basis}}$ denotes the excitation energy computed with the specified method and basis set.
The average absolute deviations from the parent methods are \SI{0.06}{eV} for CC4 and \SI{0.03}{eV} for CCSDTQ.
The deviations with respect to the corresponding ACVTZ results are consistently positive, suggesting that the composite scheme tends to systematically overestimate the corresponding large-basis results.
Moreover, Table~\ref{tab:tab1} shows that the MAE of the composite CCSDTQ with respect to the CVS-FCI reference values is slightly improved when using the composite scheme, due to a fortuitous cancellation of errors.
In the remainder of this work, only the composite scheme will be considered for CC4 and CCSDTQ, as full ACVTZ calculations for larger systems are computationally prohibitive due to their high memory requirements.

\subsection{Reference values}

\begin{table}
  \caption{Reference values of 34 core IPs (in \si{\eV}) in the ACVDZ, ACVTZ, and ACVQZ basis sets computed at the CVS-FCI level.
	The extrapolation errors associated with the CVS-FCI values are reported in parentheses.
	Some selected experimental values are also reported.}
  \label{tab:tab2}
  \begin{ruledtabular}
    \begin{tabular}{lcccc}
      & \mc{3}{c}{Basis} &  \\
      \cline{2-4}
      \mcc{} & \mcc{ACVDZ} & \mcc{ACVTZ} & \mcc{ACVQZ} & Exp. \\
      \hline
      H\textbf{F$^*$}       & 696.43 & 694.10\zz & 693.90\zz & 694.23\cite{Liu_2019}   \\
      H$_2$\textbf{O$^*$}   & 541.87 & 539.87\zz & 539.70\zz & 539.70\cite{Golze_2020} \\
      \textbf{N$^*$}H$_3$   & 407.27 & 405.63\zz & 405.49\zz & 405.52\cite{Golze_2020} \\
      \textbf{C$^*$}H$_4$   & 292.08 & 290.85\zz & 290.71\zz & 290.84\cite{Golze_2020} \\
      B\textbf{F$^*$}       & 697.63 & 695.43\zz & 695.26\zz &                         \\
      C\textbf{O$^*$}       & 544.42 & 542.47\zz & 542.32\zz & 542.10\cite{Golze_2020} \\
      \textbf{C$^*$}O       & 297.60 & 296.26\zz & 296.18\zz & 296.23\cite{Golze_2020} \\
      \textbf{N$^*_2$}      & 411.60 & 409.88\zz & 409.77\zz & 409.93\cite{Golze_2020} \\
      \textbf{F$_2^*$}      & 698.73 & 696.41\zz & 696.21(1) & 696.70\cite{PueyoBellafont_2016a} \\
      HC\textbf{N$^*$}      & 408.45 & 406.77\zz & 406.66\zz & 406.80\cite{Golze_2020} \\
      H\textbf{C$^*$}N      & 294.80 & 293.40\zz & 293.30\zz & 239.50\cite{Golze_2020} \\
      HN\textbf{O$^*$}      & 543.56 & 541.60\zz & 541.45\zz &                         \\
      H\textbf{N$^*$}O      & 411.46 & 409.84\zz & 409.75\zz &                         \\
      HO\textbf{F$^*$}      & 696.47 & 694.17\zz & 693.96\zz &                         \\
      H\textbf{O$^*$}F      & 544.50 & 542.55\zz & 542.37\zz &                         \\
      NH$_2 $\textbf{F$^*$} & 695.27 & 692.97\zz & 692.78\zz &                         \\
      \textbf{N$^*$}H$_2$F  & 410.04 & 408.44\zz & 408.34\zz &                         \\
      CH$_3$\textbf{F$^*$}  & 694.89 & 692.57\zz & 692.40(1) & 692.40\cite{Golze_2020} \\   
      \textbf{C$^*$}H$_3$F  & 294.72 & 293.56\zz & 293.45(1) & 293.56\cite{Golze_2020} \\   
      CH$_2$\textbf{O$^*$}  & 541.36 & 539.37\zz & 539.22\zz & 539.33\cite{Golze_2020} \\   
      \textbf{C$^*$}H$_2$O  & 295.68 & 294.46\zz & 294.36\zz & 294.38\cite{Golze_2020} \\   
      CH$_2$\textbf{N$^*$}H & 407.22 & 405.59(1) & 405.21(1) &                         \\   
      \textbf{C$^*$}H$_2$NH & 293.82 & 292.58(1) & 292.24(1) &                         \\   
      \textbf{C$^*_2$}H$_2$ & 292.72 & 291.28\zz & 291.17\zz & 291.25\cite{Golze_2020} \\  
      Si\textbf{O$^*$}      & 539.27 & 537.20\zz & 537.03(1) &                         \\  
      P\textbf{N$^*$}       & 407.43 & 405.74(1) & 405.61(1) &                         \\ 
      \textbf{C$^*$}S       & 294.74 & 293.39(1) & 293.34(1) &                         \\ 
      \hline                                                                         
      H\textbf{Cl$^*$}      & 2823.70 & 2821.42\zz & 2821.43\zz & 2829.8\cite{Zheng_2022} \\
      H$_2$\textbf{S$^*$}   & 2473.90 & 2471.69\zz & 2471.69(1) & 2478.5\cite{Zheng_2022} \\
      \textbf{P$^*$}H$_3$   & 2147.98 & 2145.73(1) & 2146.40(1) & 2150.9\cite{Zheng_2022} \\
      \textbf{Si$^*$}H$_4$  & 1845.50 & 1843.30\zz & 1843.79(1) & 1847.0\cite{Zheng_2022} \\
      C\textbf{S$^*$}       & 2475.05 & 2472.81\zz & 2472.88(1) &                         \\
      \textbf{P$^*$}N       & 2149.76 & 2147.47(1) & 2147.57(1) &                         \\
      \textbf{Si$^*$}O      & 1847.77 & 1845.30\zz & 1845.56(1) &                         \\
    \end{tabular}
  \end{ruledtabular}
\end{table}

\begin{table*}
  \caption{Reference values of 50 core IPs (in \si{\eV}) in the ACVDZ and ACVTZ basis sets computed at the CVS-FCI level.
	The extrapolation errors associated with the CVS-FCI values are reported in parentheses.
  Some selected experimental values are also reported.}
  \label{tab:tab3}
  \begin{ruledtabular}
    \begin{tabular}{lccclccc}
      & \mc{2}{c}{Basis} &  &  & \mc{2}{c}{Basis}  \\
      \cline{2-3} \cline{6-7}
      \mcc{} & \mcc{ACVDZ} & \mcc{ACVTZ} & Exp. & \mcc{} & \mcc{ACVDZ} & \mcc{ACVTZ} & Exp. \\
      \hline
      \textbf{C$^*_2$}H$_4$ & 292.12\zz & 290.82\zz & 290.82\cite{Golze_2020} & CH$_3$C\textbf{N$^*$}    & 407.28\zz & 405.65(1) & 405.64\cite{Liu_2019} \\
      CH$_3$\textbf{O$^*$}H & 540.97\zz & 539.04\zz & 538.88\cite{Golze_2020} & CH$_3$\textbf{C$^*$}N    & 294.08\zz & 292.79(1) & 292.45\cite{Liu_2019} \\
      \textbf{C$^*$}H$_3$OH & 293.62\zz & 292.50\zz & 292.30\cite{Golze_2020} & \textbf{C$^*$}H$_3$CN    & 293.87\zz & 292.77(1) & 292.98\cite{Liu_2019} \\
      C\textbf{O$^*_2$}     & 543.14\zz & 541.24\zz & 541.32\cite{Golze_2020} & CH$_3$\textbf{N$^*$}C    & 408.27\zz & 406.71(1) & 406.67\cite{Liu_2019} \\
      \textbf{C$^*$}O$_2$   & 299.04\zz & 297.71\zz & 297.70\cite{Golze_2020} & \textbf{C$^*$}H$_3$NC    & 294.50\zz & 293.36(1) & 293.35\cite{Liu_2019} \\
      NN\textbf{O$^*$}      & 543.36\zz & 541.44\zz & 541.42\cite{Liu_2019}   & CH$_3$N\textbf{C$^*$}    & 293.69\zz & 292.33(1) & 292.37\cite{Liu_2019} \\
      N\textbf{N$^*$}O      & 414.22\zz & 412.60\zz & 412.59\cite{Liu_2019}   & NH$_2$CH\textbf{O$^*$}   & 539.48\zz & 537.56(1) & 537.74\cite{PueyoBellafont_2016} \\
      \textbf{N$^*$}NO      & 410.39\zz & 408.70(1) & 408.71\cite{Liu_2019}   & \textbf{N$^*$}H$_2$CHO   & 408.02\zz & 406.53(2) & 406.39\cite{PueyoBellafont_2016} \\
      \textbf{F$^*$}CHO     & 696.17\zz & 693.94(1) & 694.7\cite{Ishii_1987}  & NH$_2$\textbf{C$^*$}HO   & 295.63\zz & 294.48(1) & 294.95\cite{PueyoBellafont_2016} \\
      FCH\textbf{O$^*$}     & 542.05\zz & 540.10(1) & 540.1\cite{Ishii_1987}  & CH$_3$N\textbf{O$^*$}    & 542.33(1) & 540.42(2) &  \\
      F\textbf{C$^*$}HO     & 298.24\zz & 297.03(1) & 297.1\cite{Ishii_1987}  & CH$_3$\textbf{N$^*$}O    & 410.65(1) & 409.12(2) &  \\
      HNC\textbf{O$^*$}     & 542.06\zz & 540.14(1) & 540.16\cite{Jolly_1984} & \textbf{C$^*$}H$_3$NO    & 293.12\zz & 292.02(1) &  \\
      H\textbf{N$^*$}CO     & 408.01\zz & 406.40(1) & 406.44\cite{Jolly_1984} & CH$_2$CH\textbf{F$^*$}   & 695.37(1) & 693.09(1) & 693.26\cite{Jolly_1984} \\
      HN\textbf{C$^*$}O     & 297.30\zz & 295.94(1) & 295.89\cite{Jolly_1984} & CH$_2$\textbf{C$^*$}HF   & 294.69\zz & 293.44(1) & 293.48\cite{Jolly_1984} \\
      HCO\textbf{O$^*$}H    & 542.52(1) & 540.68(1) & 540.69\cite{Golze_2020} & \textbf{C$^*$}H$_2$CHF   & 292.33(1) & 291.03(1) & 291.10\cite{Jolly_1984} \\
      HC\textbf{O$^*$}OH    & 540.81(1) & 538.87(2) & 539.02\cite{Golze_2020} & CH$_3$CH\textbf{O$^*$}   & 540.52\zz & 538.48(1) & 538.50\cite{PueyoBellafont_2016a} \\
      H\textbf{C$^*$}OOH    & 296.93(1) & 295.68(2) & 295.75\cite{Golze_2020} & CH$_3$\textbf{C$^*$}HO   & 295.19\zz & 293.97(1) & 294.00\cite{PueyoBellafont_2016a} \\
      H$_2$C\textbf{N$^*$}N & 411.12\zz & 409.43(1) &                         & \textbf{C$^*$}H$_3$CHO   & 292.58\zz & 291.36(1) & 291.60\cite{PueyoBellafont_2016a} \\
      H$_2$CN\textbf{N$^*$} & 408.88\zz & 407.12(1) &                         & \textbf{N$^*$}$_2$CH$_2$ & 409.66\zz & 407.99(1) &  \\
      H$_2$\textbf{C$^*$}NN & 292.95\zz & 291.66(1) &                         & N$_2$\textbf{C$^*$}H$_2$ & 294.08\zz & 292.92\zz &  \\
      \cline{5-8}
      H$_2$CC\textbf{O$^*$} & 542.13\zz & 540.17(1) & 540.25\cite{Jolly_1984} & \textbf{P$^*$}$_2$       & 2147.92(1) & 2145.61(1) &  \\
      H$_2$C\textbf{C$^*$}O & 296.10\zz & 294.74(1) & 294.73\cite{Jolly_1984} & \textbf{Cl$^*$}$_2$      & 2824.17(1) & 2822.27(3) & 2830.2\cite{Zheng_2022} \\ 
      H$_2$\textbf{C$^*$}CO & 292.41\zz & 291.15(1) & 291.13\cite{Jolly_1984} & CH$_2$\textbf{S$^*$}     & 2473.35\zz & 2471.11(1) &  \\
      \textbf{C$^*$}H$_2$S  & 293.57\zz & 292.28\zz &                         & OC\textbf{S$^*$}         & 2474.28(1) & 2472.09(2) & 2478.7\cite{Zheng_2022} \\ 
      \textbf{O$^*$}CS      & 542.18(1) & 540.25(2) & 540.3\cite{Jolly_1984}  &  &  &  &  \\
      O\textbf{C$^*$}S      & 296.72(1) & 295.37(3) & 295.2\cite{Jolly_1984}  &  &  &  &  \\
    \end{tabular}                                
  \end{ruledtabular}
\end{table*}

Table~\ref{tab:tab2} reports 27 second-row and 7 third-row CVS-FCI core IPs computed using the ACVDZ, ACVTZ, and ACVQZ basis sets.
Table~\ref{tab:tab3} presents an additional 46 second-row and 4 third-row core IPs for slightly larger systems, for which only CVS-FCI calculations with the ACVDZ and ACVTZ basis sets could be performed.

For degenerate core orbitals (\ce{N2}, \ce{F2}, \ce{C2H4}, etc), only a single value is reported.
They have been obtained using the above procedure for CVS-FCI on top of a spatially symmetry-broken determinant.
\titou{These symmetry-broken determinants have been obtained by mixing the $1\sigma_g$ and $1\sigma_u$ orbitals before performing the MOM calculations.}
At the mean-field level, it is known that breaking spatial symmetry greatly improves the energy with respect to experiment.\cite{Bagus_1972,Cederbaum_1977a}
Upon inclusion of correlation, one should recover the correct result even with symmetry-preserved orbitals.\cite{Norman_2018}
However, because CVS-FCI is not invariant with respect to orbital rotations, here a symmetry-broken mean-field starting point is considered at the correlated level.
See also Ref.~\onlinecite{Schoffler_2008} for an experimental perspective on the localization of core holes in \ce{N2}.

In what follows, the CVS-FCI (with natural orbitals, see Sec.~\ref{sec:small-molecules}) results in the ACVTZ basis set are used as reference values to assess the performance of approximate methods.
However, it is important to emphasize that these values are not fully converged with respect to basis set size.
As seen in Table~\ref{tab:tab2}, the average shift in core IPs when moving from triple- to quadruple-zeta basis sets is still on the order of \SIrange{0.1}{0.2}{eV}.

For comparison, selected experimental values are also reported in Tables~\ref{tab:tab2} and~\ref{tab:tab3}.
The average absolute deviation of CVS-FCI (in the ACVTZ basis) from experiment is \SI{0.11}{eV} for second-row elements.
This good agreement is largely the result of fortuitous error cancellation, since neither relativistic effects nor basis set incompleteness corrections are accounted for.
\titou{Indeed, relativistic corrections have been estimated to contribute approximately \SI{0.12}{eV} for carbon, \SI{0.24}{eV} for nitrogen, \SI{0.42}{eV} for oxygen, and \SI{0.72}{eV} for fluorine, by Golze and co-workers.\cite{Golze_2020}
If these corrections are added to CVS-FCI, the MAE with respect to experiment becomes \SI{0.28}{eV} and the associated MSE is \SI{0.26}{eV}.
This systematic overestimation is expected to diminish upon enlargement of the basis set.}

In contrast, the mean absolute deviation of CVS-FCI (in the ACVTZ basis) from experiment for third-row atoms is substantially larger, \SI{6.43}{eV}, highlighting the much greater importance of relativistic effects in this case.
It is well established that scalar relativistic corrections are insufficient for third-row elements; spin-orbit coupling and possibly higher-order effects must be explicitly included for reliable comparison with experimental data.\cite{Niskanen_2011,Zheng_2022}

\begin{figure*}
  \includegraphics[width=0.3\linewidth]{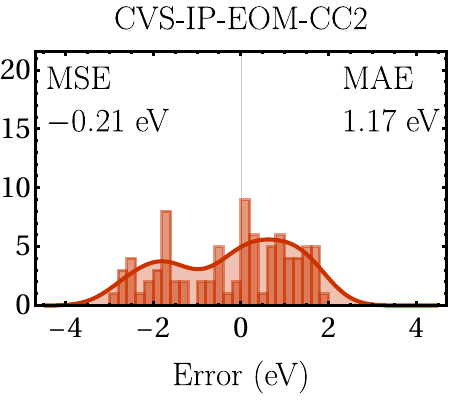}
  \hspace{0.03\linewidth}
  \includegraphics[width=0.3\linewidth]{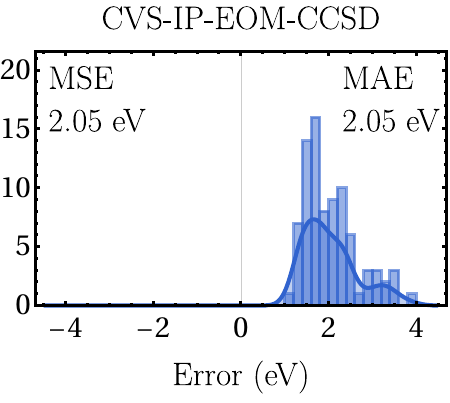}
  \hspace{0.03\linewidth}
  \includegraphics[width=0.3\linewidth]{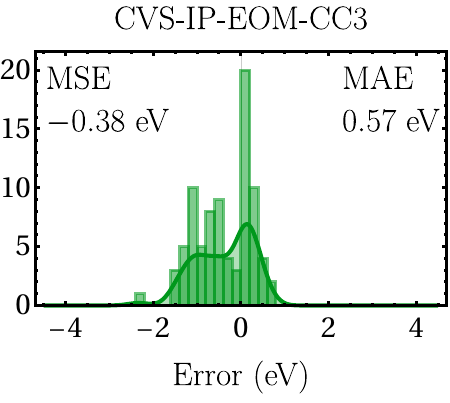}
  \\
  \vspace{0.03\linewidth}
  \includegraphics[width=0.3\linewidth]{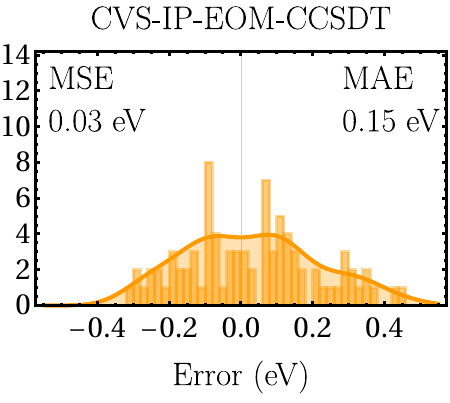}
  \hspace{0.03\linewidth}
  \includegraphics[width=0.3\linewidth]{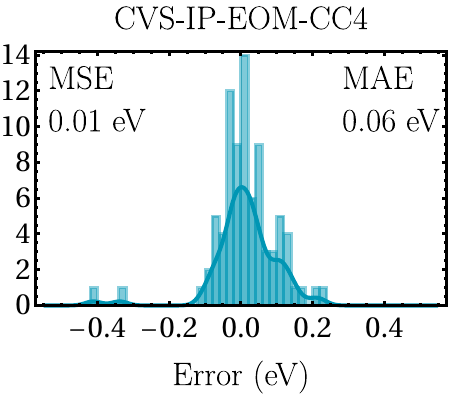}
  \hspace{0.03\linewidth}
  \includegraphics[width=0.3\linewidth]{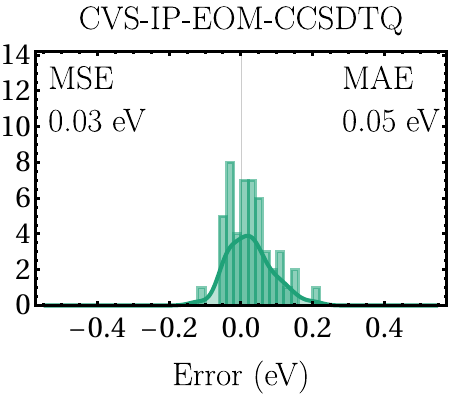}
  \caption{\titou{Histogram of the errors (with respect to CVS-FCI) for 84 core IPs in the ACVTZ basis set calculated using CVS-IP-EOM-CC2, -CCSD, -CC3, -CSDT, -CC4, and -CCSDTQ (the latter available for only 52 cases). Results for CC4 and CCSDTQ were obtained using a composite basis set scheme involving CCSDT and the ACVDZ basis set [see Eq.~\eqref{eq:composite}]. Note the change in the horizontal-axis scale between the top and bottom rows.}}
  \label{fig:fig1}
\end{figure*}
  

\subsection{Approximate methods}

Now that the reference values have been defined, let us gauge the performance of some approximate wave function methods.
\titou{Figure~\ref{fig:fig1}} shows the histogram of errors (with respect to CVS-FCI) of six variants of CC.
The corresponding raw data can be found in the \SupInf, while the associated statistical descriptors are provided in Table~\ref{tab:tab4}.
Every method considered here uses the same ACVTZ basis set as the reference values, except for CC4 and CCSDTQ, which use the composite basis set scheme described in Sec.~\ref{sec:small-molecules}.
Despite the use of a composite basis set scheme, memory limitations restricted the CCSDTQ calculations to only 52 core IPs

First, one can observe that the systematic improvability of CC is once again verified as the MAE decreases from \SI{2.05}{\eV} to \SI{0.15}{\eV} upon inclusion of triple excitations.
Taking into account quadruple excitations further improves these results with a MAE of \SI{0.05}{\eV}.
However, there are still a few outliers with errors larger than \SI{0.1}{\eV} even at the CCSDTQ level, highlighting the difficulty in recovering relaxation effects in a linear-response framework.
Note that the CCSDTQ MAE is of the same order of magnitude as the error introduced by the composite basis set scheme (see Sec.~\ref{sec:small-molecules}).
Hence, one should be cautious about this result as the CCSDTQ MAE in the ACVTZ basis might be slightly larger.
The MSEs of CCSDT (\SI{0.03}{\eV}) and CCSDTQ (\SI{0.03}{\eV}) show that the error distributions are nicely centered around zero.
On the other hand, CCSD systematically overestimates the reference values with a MSE of \SI{2.05}{\eV}.

\begin{table*}
  \caption{Statistical descriptors (in \si{\eV}) associated with the errors (with respect to CVS-FCI) for 84 core IPs in the ACVTZ basis set calculated using CVS-IP-EOM-CC2, -CCSD, -CC3, -CCSDT, -CC4, -CCSDTQ, $\Delta$ROHF, $\Delta$UHF, $\Delta$UMP2, and $G_0W_0$@PBEh($\alpha$).
    The results for CVS-IP-EOM-CC4 and -CCSDTQ have been obtained using a composite basis set scheme using CVS-IP-EOM-CCSDT and the ACVDZ basis set [see Eq.~\eqref{eq:composite}].
    The results for $G_0W_0$@PBEh($\alpha$) have been obtained using $\alpha=0.45$ and $\alpha=0.70$ for second- and third-row atoms, respectively.\label{tab:tab4}}
  \begin{ruledtabular}
    \begin{tabular}{lcccccccccc}
           & \mcc{CC2} & \mcc{CCSD} & \mcc{CC3} & \mcc{CCSDT} & \mcc{CC4} & \mcc{CCSDTQ} & \mcc{$\Delta$ROHF} & \mcc{$\Delta$UHF} & \mcc{$\Delta$UMP2} & \mcc{$G_0W_0$@PBEh($\alpha$)} \\
      \hline
      MSE  &  -0.21 & 2.05 &  -0.38 & \m0.03 & \m0.01 & \m0.03 &  -0.05 &  -0.50 & \m0.66 &  -0.02 \\
      MAE  & \m1.17 & 2.05 & \m0.57 & \m0.15 & \m0.06 & \m0.05 & \m0.57 & \m0.74 & \m0.67 & \m0.48 \\
      RMSE & \m1.40 & 2.14 & \m0.74 & \m0.19 & \m0.09 & \m0.07 & \m0.73 & \m0.88 & \m0.87 & \m0.58 \\
      SDE  & \m1.39 & 0.63 & \m0.64 & \m0.18 & \m0.09 & \m0.06 & \m0.73 & \m0.73 & \m0.57 & \m0.59 \\
      Min  &  -2.95 & 1.14 &  -2.33 &  -0.30 &  -0.41 &  -0.12 &  -1.64 &  -1.96 &  -0.29 &  -1.66 \\
      Max  & \m1.88 & 3.86 & \m0.62 & \m0.46 & \m0.22 & \m0.20 & \m1.81 & \m1.48 & \m3.15 & \m1.26 \\
    \end{tabular}
  \end{ruledtabular}
\end{table*}

The accuracy of the popular family of CC$n$ approximations can also be assessed.
It can be seen in Table~\ref{tab:tab4} that CC4 does not deteriorate the performance of CCSDTQ, while CC3 seems to underestimate CCSDT on average.
This leads to a poorer MAE of \SI{0.57}{\eV}.
The CCSD results are also underestimated by CC2, but because CCSD systematically overestimates the reference values, this results in an improved MAE (\SI{1.17}{\eV}) for CC2 with respect to CCSD.

\begin{figure*}
  \includegraphics[width=0.24\linewidth]{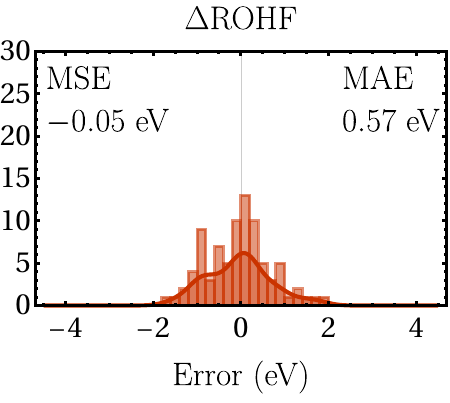}
  \includegraphics[width=0.24\linewidth]{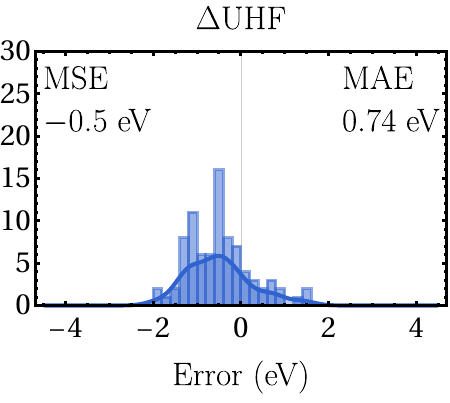}
  \includegraphics[width=0.24\linewidth]{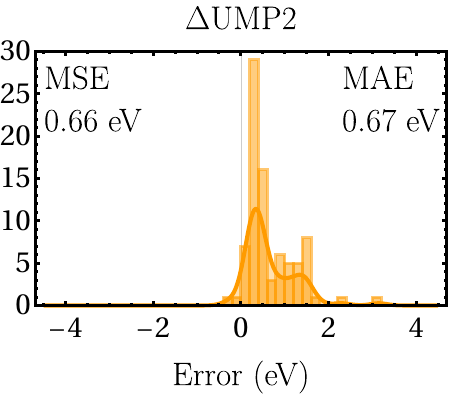}
  \includegraphics[width=0.24\linewidth]{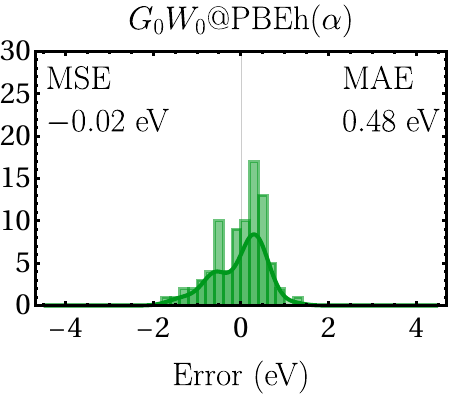}
  \caption{Histogram of the errors (with respect to CVS-FCI) for 84 core IPs in the ACVTZ basis set calculated using $\Delta$ROHF, $\Delta$UHF, $\Delta$UMP2, and $G_0W_0$@PBEh($\alpha$). The results for the latter have been obtained using $\alpha=0.45$ and $\alpha=0.70$ for second- and third-row atoms, respectively. \label{fig:fig3}}
\end{figure*}

Figure~\ref{fig:fig3} presents the error distribution of three state-specific methods: $\Delta$ROHF, $\Delta$UHF, and $\Delta$UMP2.
The non-Aufbau ROHF determinants correspond to the starting points of the SCI procedure as described in Sec.~\ref{sec:sCI}.
The $\Delta$ROHF method performs as well as CC3 with a MAE of \SI{0.57}{\eV} (see Table~\ref{tab:tab4}), while being more centered around zero with its MSE of \SI{-0.05}{\eV}.
Note that the IPs corresponding to degenerate orbitals were obtained using spatially symmetry-broken orbitals.
In order to gauge the performance of state-specific correlated methods, we decided to use an unrestricted reference starting point to avoid the dependence of the MP2 correlation energy on the choice of the ROHF Fock matrix.
As shown in Table~\ref{tab:tab4}, $\Delta$UHF yields a larger MAE (\SI{0.74}{\eV}) than $\Delta$ROHF, primarily due to an average underestimation of core IPs, as indicated by its MSE of \SI{-0.50}{\eV}. 
While $\Delta$UMP2 corrects this systematic bias, it slightly overcompensates, resulting in only a modest improvement in MAE (\SI{0.67}{\eV}).
We also observe excellent performance for $\Delta$UCCSD, comparable to CVS-IP-EOM-CCSDT and significantly improved over $\Delta$UMP2. 
However, these state-specific optimizations could not be converged for the majority of the ionized states of this set and are therefore not included in this study.
While CVS-like approximations can also be devised to alleviate these convergence issues, such developments lie beyond the scope of this work.\cite{Zheng_2019,Zheng_2022}

Finally, the performance of the $G_0W_0$ method is gauged.
For second-row elements, the calculations were performed starting from orbitals and energies obtained with a customized PBE hybrid functional containing 45\% exact exchange [PBEh(45)].
This functional, which incorporates a large fraction of non-local exchange, has been shown to perform particularly well for core binding energies in both solids and molecules.~\cite{Golze_2020,Zhu_2021,Vo_2025}
In contrast, for third-row elements, this starting point fails to yield well-defined quasiparticle solutions: all graphical solutions of the quasiparticle equation exhibit spectral weights smaller than 0.5. 
However, following the methodology of Ref.~\onlinecite{Golze_2020}, we have found that employing a PBEh(70) starting point yields well-defined quasiparticle solutions for all third-row core IPs considered in this study.
This alternative starting point yields a MAE of \SI{0.51}{\eV} for the 11 third-row core IPs while $G_0W_0$@PBEh(45) has a MAE of \SI{0.48}{\eV} for the 73 second-row core IPs.

\begin{table}[t!]
  \caption{Atom-specific MSEs and MAEs (with respect to CVS-FCI), in \si{\eV}, for carbon, nitrogen, oxygen, and fluorine in the ACVTZ basis set calculated using CVS-IP-EOM-CC2, -CCSD, -CC3, -CCSDT, -CC4, -CCSDTQ, $\Delta$ROHF, $\Delta$UHF, $\Delta$UMP2, and $G_0W_0$@PBEh($\alpha$).
    The results for CVS-IP-EOM-CC4 and -CCSDTQ have been obtained using a composite basis set scheme using CVS-IP-EOM-CCSDT and the ACVDZ basis set [see Eq.~\eqref{eq:composite}].
    The results for $G_0W_0$@PBEh($\alpha$) have been obtained using $\alpha=0.45$ for these second-row atoms.}
  \label{tab:tab5}
  \begin{ruledtabular}
    \begin{tabular}{lcccccccc}
      & \mc{4}{c}{MSE} & \mc{4}{c}{MAE} \\
      \cline{2-5} \cline{6-9}
      & \ce{C} & \ce{N} & \ce{O} & \ce{F} & \ce{C} & \ce{N} & \ce{O} & \ce{F} \\
      \hline
      CC2          & \m1.23 &  -0.27 &  -1.74 &  -2.52 & 1.23 & 0.46 & 1.74 & 2.52 \\
      CCSD         & \m1.56 & \m1.83 & \m2.22 & \m2.24 & 1.56 & 1.83 & 2.22 & 2.24 \\
      CC3          & \m0.23 &  -0.15 &  -1.07 &  -1.32 & 0.24 & 0.25 & 1.07 & 1.32 \\
      CCSDT        & \m0.14 & \m0.05 &  -0.12 &  -0.21 & 0.15 & 0.14 & 0.14 & 0.21 \\
      CC4          & \m0.07 & \m0.03 &  -0.04 &  -0.02 & 0.08 & 0.05 & 0.05 & 0.04 \\
      CCSDTQ       & \m0.08 & \m0.04 &  -0.03 &  -0.02 & 0.08 & 0.05 & 0.04 & 0.03 \\
      $\Delta$ROHF & \m0.47 & \m0.02 &  -0.76 &  -1.13 & 0.49 & 0.32 & 0.76 & 1.13 \\
      $\Delta$UHF  &  -0.04 &  -0.60 &  -1.10 &  -1.45 & 0.52 & 0.68 & 1.10 & 1.45 \\
      $\Delta$UMP2 & \m0.74 & \m0.90 & \m0.46 & \m0.42 & 0.75 & 0.90 & 0.46 & 0.42 \\
      $G_0W_0$     & \m0.35 & \m0.10 & -0.54 & -1.19 & 0.42 & 0.20 & 0.54 & 1.19 \\
    \end{tabular}
  \end{ruledtabular}
\end{table}

To conclude this section, the MSEs and MAEs of the ten approximate methods considered here are computed for each atom of the second row separately.
The corresponding values are reported in Table~\ref{tab:tab5}.
It shows that the MSEs are decreasing when going from carbon to fluorine for every method except CCSD and $\Delta$UMP2.
Regarding the absolute performance, the atom-specific MAEs show that CC3 performs much better for carbon and nitrogen than for oxygen and fluorine.
Overall, fluorine core IPs have the largest MAE for every method except $\Delta$UMP2, CC4, and CCSDTQ.
This is probably due to the use of the composite basis set scheme for the latter two.
Finally, one should remember that fluorine core IPs are also the largest in absolute value, which also partly explains this trend for atom-specific MAEs.

\section{Conclusion}
\label{sec:conclusion}

In this work, we have established the first large-scale benchmark of non-relativistic core IPs computed at the CVS-FCI level.
The dataset comprises 84 values for small molecules containing second- and third-row atoms, consistently reported in the ACVTZ basis and extended to ACVQZ whenever feasible.
These results define chemically accurate TBEs within fixed finite-basis sets, providing a reference that isolates correlation and relaxation effects from other physical contributions, such as relativistic corrections, vibrational structure, and continuum coupling.

Using these references, we assessed the performance of widely used approximate methods, including the hierarchy of EOM-CC approaches, as well as some state-specific methods and one $GW$ variant.
The results demonstrate the systematic improvability of CC theory: errors decrease from several \si{\eV} at the CCSD level to below \SI{0.1}{\eV} upon inclusion of quadruple excitations.
Approximate schemes such as CC3 and CC4 already reach useful accuracy, albeit with slightly larger deviations, while $\Delta$SCF provides a cost-effective alternative with errors on par with CC3.
Nevertheless, even the most advanced correlation treatments occasionally deviate by \SI{0.1}{\eV}, highlighting the inherent challenges of fully capturing relaxation within a linear-response framework \titou{that starts from neutral ground-state orbitals}.
\titou{On the other hand, state-specific methods offer a way to recover this orbital relaxation, but they are hindered by convergence issues once correlation is included.}
It was shown that the good performance of $G_0W_0$@PBEh(45) for second-row core IPs does not extend to third-row core IPs.
Instead, a PBEh(70) starting point must be used to achieve the same accuracy, which calls for a study of the performance of self-consistent, either fully or partially, $GW$ schemes.

Beyond serving as a benchmark, the present dataset lays the groundwork for systematic development and calibration of new electronic structure methods for core-level spectroscopy.
It extends the scope of the \textsc{quest} database\cite{Loos_2025} from neutral to charged core excitations,\cite{Marie_2024} offering a unified and internally consistent set of theory-based references for both valence and core processes.
These data should prove valuable for validating approximate methods, guiding basis set and methodological choices, and enabling more predictive modeling of XPS spectra in molecules.
\titou{These new accurate reference values could also be used to train machine-learning models, as this field has shown promising results in the prediction of core-electron binding energies.\cite{Sun_2022,Golze_2022,Zarrouk_2024,Tripathy_2024,Porcelli_2025,Fouda_2026}}

Future work will focus on broadening this foundation.
A natural extension is the computation of transition intensities, essential for direct comparison with experimental spectra.
While not straightforward within the present state-specific SCI framework, intensity calculations are feasible, as shown in recent studies,\cite{Mejuto-Zaera_2021,Ferte_2020,Ferte_2022} and will be explored in forthcoming work.
Another important perspective is the inclusion of satellite features in core IPs, which encode many-body shake-up processes and provide further insight into correlation effects beyond principal ionizations.\cite{Marie_2024,Loos_2024b,Kocklauner_2025a,Kocklauner_2025b}
We plan to report on this additional layer of reference data in the near future.

Altogether, we expect that this benchmark will stimulate the development of novel approximate methods for core spectroscopy, building on the success of high-accuracy reference datasets in advancing quantum chemical approaches to excited and ionized states.

\section*{Supporting Information}
\label{sec:supmat}

See the \SupInf for the ground-state geometry of each system considered in this study, and the core IPs computed at various levels of theory \titou{(raw data associated with Fig.~\ref{fig:fig1})}.

\acknowledgments{The author would like to thank Anthony Scemama for technical assistance. 
This project has received funding from the European Research Council (ERC) under the European Union's Horizon 2020 research and innovation programme (Grant agreement No.~863481).
It is also funded, in part, by the Agence Nationale de la Recherche (ANR), grant ANR-25-CE29-4996.
This work used the HPC resources from CALMIP (Toulouse) under allocation 2025-18005 and 2026-18005.}


\section*{References}

\bibliography{coreIP.bib}

\end{document}